\documentclass[aps,pra,reprint,showpacs,superscriptaddress,twocolumn]{revtex4-1}%
\usepackage{amsfonts}
\usepackage{mathrsfs}
\usepackage{amsmath}
\usepackage{amssymb}
\usepackage{graphicx}
\usepackage{color}%
\usepackage{mathtools}
\usepackage{booktabs}
\usepackage[colorlinks,linkcolor=blue,citecolor=blue,hyperindex,bookmarks=false,pdfstartview=FitH]{hyperref}

\providecommand{\U}[1]{\protect\rule{.1in}{.1in}}

\begin{document}
\title{Single-photon frequency conversion via a giant $\Lambda$-type atom}

\author{Lei Du}
\address{Beijing Computational Science Research Center, Beijing 100193, China}
\author{Yong Li}
\email{liyong@csrc.ac.cn}
\address{Beijing Computational Science Research Center, Beijing 100193, China}
\address{Synergetic Innovation Center for Quantum Effects and Applications, Hunan Normal University, Changsha 410081, China}

\date{\today }

\begin{abstract}
We study single-photon scattering via a giant $\Lambda$-type atom, where both atomic transitions are coupled with the modes of a single waveguide at two separated points. The giant-atom structure introduces phase-dependent interference effects to both elastic (frequency-preserving) and inelastic (frequency-converting) scattering processes, which modify the corresponding decay rates (as well as the transition frequencies) such that the giant atom is capable of accessing the various limits of a small one. The condition of the optimal frequency conversion is also identified and shown to be phase dependent. Moreover, we consider the combination of the giant-atom interference and the Sagnac quantum interference by further inserting a Sagnac interferometer at each of the coupling points. It is shown that the two kinds of interference effects are compatible and play independent roles, such that efficient frequency conversion with unit efficiency can be achieved in addition to the phase-dependent phenomena induced by the giant-atom structure.
\end{abstract}

%\pacs{42.50.-p, 42.50.Pq, 42.65.-k}
\maketitle

%\tableofcontents
\section{Introduction}\label{sec1}
Waveguide quantum electrodynamics (QED) is an emerging field that studies the interactions between atoms and electromagnetic fields confined in one-dimensional open waveguides. It provides an alternative platform for enhancing light-matter interactions, where the electromagnetic modes can interact strongly with atoms due to the transverse confinement of the waveguides~\cite{Wreview1,Wreview2}. Different from cavity QED systems, waveguides support in general a continuum of modes such that the bandwidth limitation that is typical for cavities (due to their finite linewidths) can be greatly relaxed~\cite{Wreview1}. Moreover, long-range interactions between remote atoms (resonators), which are vital for studying many-body physics and realizing large-scale quantum networks, can be mediated by the traveling photons in waveguides~\cite{fan1999,xiao2008,xiao2010,JPan,BBLi,Solano}. To date, several candidates of waveguide QED systems have been developed, such as trapped (natural) atoms coupled with optical fibers~\cite{fiber1,fiber2,fiber3} or photonic crystal waveguides~\cite{crystal1,crystal2,crystal3} and superconducting qubits coupled with transmission lines~\cite{tline1,tline2,tline3,tline4,tline5}. The technological improvements and growing research interests have led to progress in waveguide QED, e.g.,
chiral photon-atom interactions~\cite{chiral1,chiral2,chiral3}, single-photon routers~\cite{rout1,rout2,rout3,rout4}, topologically induced unconventional quantum optics~\cite{tuqo,PRXtopo}, and entangled-state preparations~\cite{entpre1,entpre2,entpre3} to name a few. In particular, frequency conversion at the single-photon level can be achieved with a three-level atom (either V type or $\Lambda$ type) coupled to a single waveguide, where the nonlinear optical process can be accomplished with only a single photon~\cite{efficient,shenFC,njpthree,LSzhw}.

Considering that the sizes of atoms (either natural or artificial) are in general much smaller than the wavelengths of the waveguide modes, dipole approximation is usually adopted to regard the atoms as single points~\cite{walls}. Recent experimental progress revealed that the approximation should be modified when atoms interact with the waveguide modes at multiple points that are separated by large distances. For example, transmons are able to interact with surface acoustic waves via interdigital transducers~\cite{transmon1,transmon2,transmon3}, with the separations between different transmon-waveguide coupling points comparable to or even much larger than the typical wavelengths ($\sim10^{-6}\,\textrm{m}$) of the surface acoustic waves therein. Alternatively, one can couple single atoms with a bent waveguide so that they can interact many times with the waveguide modes, such as the Xmon version demonstrated in Refs.~\cite{Lamb,braided,engineer}. Such structures, which are referred to as giant atoms, demonstrate striking interference effects that depend on both the atomic size (i.e., the separation between different coupling points) and the photonic frequency~\cite{Lamb,braided,engineer,Lamb2,GANori,GAWang,oscillate,EIAT,DLoe,tunchiral}. Moreover, non-Markovian retardation effects should also be included if the separations between different coupling points are comparable to or even larger than the coherence length of the emitted photons, with which the dynamics can markedly deviate from the Markovian predictions~\cite{Lamb2,oscillate,kockumRev,longhiGA,DLretard}. Recently, giant-atom structures have also been extended to higher dimensions by using optical lattices of cold atoms~\cite{highD}. Despite the seminal works above, investigations of single-photon frequency conversions with such giant-atom interferences are still lacking.

In this paper, we study single-photon scattering via a giant $\Lambda$-type atom, which can be excited by the waveguide modes via both transitions and is coupled with the waveguide at two separated points. The input photon can either be transmitted or reflected directly or undergo frequency conversion, depending on which of the two lower-energy states of the atom is finally occupied. From a comparison with the small-atom case~\cite{efficient,shenFC}, where the scattering behavior is only determined by the ratio of the waveguide-induced decay rates of the two transitions, we reveal that the scattering here also depends on two phase factors which are related to the two transition frequencies respectively as well as the separation between the two coupling points. The presence of two coupling points results in phase-dependent interference effects, which are in general different for the two atomic transitions. Such interference effects thus affect the scattering behaviors via modifying the transition frequencies and decay rates of the atom. By tuning the phases, the giant atom is able to demonstrate phenomena that typically occur in various limits of the small-atom case, such as perfect transmission over the whole frequency range and total reflection. Moreover, we insert a Sagnac interferometer at each of the coupling points, which results in quantum interferences between the counterpropagating waveguide modes and thereby enables efficient frequency conversion. On the other hand, all the aforementioned giant-atom effects can still be observed and the parametric conditions remain unchanged in this case, implying that the two kinds of interference effects play their roles in parallel.

\section{Models and methods}\label{sec2}
\begin{figure}[ptb]
\centering
\includegraphics[width=8.2 cm]{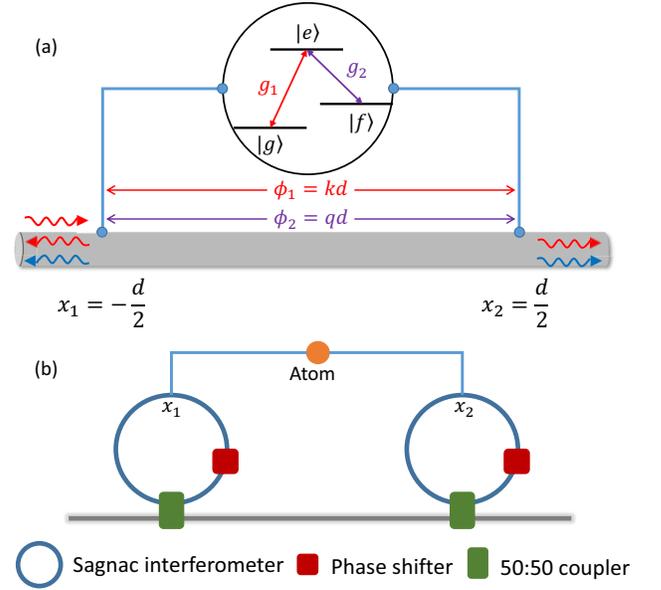}
\caption{(a) Schematic illustration of the single-photon frequency converter, where both transitions of a $\Lambda$-type three-level atom couple twice with a waveguide at $x_{1}=-d/2$ and $x_{2}=d/2$. (b) Sketch of the Sagnac-interference scheme.}\label{fig1}
\end{figure}

We consider in this paper a giant $\Lambda$-type three-level atom which couples twice with a single waveguide. As shown in Fig.~\ref{fig1}(a), both transitions $|g\rangle\leftrightarrow|e\rangle$ and $|f\rangle\leftrightarrow|e\rangle$ of the atom are coupled with the waveguide modes via two coupling points located at $x_{1}=-d/2$ and $x_{2}=d/2$, where $|g\rangle$, $|f\rangle$, and $|e\rangle$ are the ground, middle, and excited states, respectively. In the following, we assume that $|f\rangle$ is a metastable state, which can be achieved if $|g\rangle$ and $|f\rangle$ are two hyperfine levels generated by, e.g., the Zeeman effects. The real-space Hamiltonian of the system can be written as ($\hbar=1$)
\begin{equation}
\begin{split}
H&=H_{\textrm{w}}+H_{\textrm{a}}+H_{\textrm{int}},\\
H_{\textrm{w}}&=iv_{g}\int_{-\infty}^{+\infty}dx\Big[a_{L}^{\dag}(x)\frac{\partial}{\partial x}a_{L}(x)-a_{R}^{\dag}(x)\frac{\partial}{\partial x}a_{R}(x)\Big],\\
H_{\textrm{a}}&=\omega_{f}|f\rangle\langle f|+\omega_{e}|e\rangle\langle e|,\\
H_{\textrm{int}}&=\int_{-\infty}^{+\infty}dxP(x)\{g_{1}[a_{R}^{\dag}(x)+a_{L}^{\dag}(x)]|g\rangle\langle e|\\
&\quad+g_{2}[a_{R}^{\dag}(x)+a_{L}^{\dag}(x)]|f\rangle\langle e|+\text{H.c.}\},
\end{split}
\label{eq1}
\end{equation}
where $P(x)=\delta(x+d/2)+\delta(x-d/2)$; $H_{\textrm{w}}$ is the free Hamiltonian of the waveguide modes with $v_{g}$ the group velocity; $a_{R}^{\dag}$ ($a_{R}$) and $a_{L}^{\dag}$ ($a_{L}$) are the creation (annihilation) operators of the right-moving and left-moving photons in the waveguide, respectively; $H_{\textrm{a}}$ is the free Hamiltonian of the $\Lambda$-type atom, where $\omega_{f}$ and $\omega_{e}$ are the energies of states $|f\rangle$ and $|e\rangle$ with respect to the ground state, respectively; and $H_{\textrm{int}}$ describes the interactions between the atom and the waveguide, where $g_{1}$ and $g_{2}$ are the coupling strengths of transitions $|g\rangle\leftrightarrow|e\rangle$ and $|f\rangle\leftrightarrow|e\rangle$ with the waveguide modes, respectively. We have assumed that the coupling strengths are identical for both coupling points, i.e., $g_{j}(x_{1})=g_{j}(x_{2})$ ($j=1,2$), with which some limit phenomena can be achieved, as will be discussed below. In the single-photon manifold, the eigenstate of Hamiltonian (\ref{eq1}) can be written as
\begin{equation}
\begin{split}
|\psi\rangle&=\sum_{\alpha=g,f}\int_{-\infty}^{+\infty}dx[R_{\alpha}(x)a^{\dag}_{R}(x)|0,\alpha\rangle\\
&\quad+L_{\alpha}(x)a^{\dag}_{L}(x)|0,\alpha\rangle]+u_{e}|0,e\rangle,
\end{split}
\label{eq2}
\end{equation}
where $R_{\alpha}(x)$ [$L_{\alpha}(x)$] is the probability amplitude of creating a right-moving (left-moving) photon in the waveguide at position $x$ and the atom finally in state $|\alpha\rangle$. In addition, $u_{e}$ is the probability amplitude of the atom in the excited state, finding no photon in the waveguide. By solving the Schr\"{o}dinger equation $H|\psi\rangle=E|\psi\rangle$, one can obtain the following equations of the probability amplitudes
\begin{equation}
\begin{split}
ER_{g}(x)&=-iv_{g}\frac{\partial}{\partial x}R_{g}(x)+g_{1}P(x)u_{e},\\
EL_{g}(x)&=iv_{g}\frac{\partial}{\partial x}L_{g}(x)+g_{1}P(x)u_{e},\\
ER_{f}(x)&=\Big(\omega_{f}-iv_{g}\frac{\partial}{\partial x}\Big)R_{f}(x)+g_{2}P(x)u_{e},\\
EL_{f}(x)&=\Big(\omega_{f}+iv_{g}\frac{\partial}{\partial x}\Big)L_{f}(x)+g_{2}P(x)u_{e},\\
Eu_{e}&=\omega_{e}u_{e}+g_{1}P(x)[R_{g}(x)+L_{g}(x)]\\
&\quad+g_{2}P(x)[R_{f}(x)+L_{f}(x)].
\end{split}
\label{eq3}
\end{equation}

We first assume that a single photon of wave vector $k$ ($k>0$) is incident from the far left of the waveguide and the atom is initialized in the ground state $|g\rangle$. If the atom is excited by the input photon that is nearly resonant with the transition $|g\rangle\leftrightarrow|e\rangle$, it can spontaneously decay to the lower-energy state $|g\rangle$ or $|f\rangle$ and emit a photon of frequency $v_{g}k$ or $v_{g}k-\omega_{f}$~\cite{efficient,shenFC}. In the latter situation, the photon undergoes an inelastic scattering and the frequency is down-converted. Equation~(\ref{eq3}) can be solved according to the ansatz
\begin{equation}
\begin{split}
R_{g}(x)&=\Big\{\Theta\Big(-x-\frac{d}{2}\Big)+A\Big[\Theta\Big(x+\frac{d}{2}\Big)\\
&\quad-\Theta\Big(x-\frac{d}{2}\Big)\Big]+t_{1}\Theta\Big(x-\frac{d}{2}\Big)\Big\}e^{ikx},\\
L_{g}(x)&=\Big\{r_{1}\Theta\Big(-x-\frac{d}{2}\Big)+B\Big[\Theta\Big(x+\frac{d}{2}\Big)\\
&\quad-\Theta\Big(x-\frac{d}{2}\Big)\Big]\Big\}e^{-ikx},\\
R_{f}(x)&=\Big\{M\Big[\Theta\Big(x+\frac{d}{2}\Big)-\Theta\Big(x-\frac{d}{2}\Big)\Big]\\
&\quad+t_{2}\Theta\Big(x-\frac{d}{2}\Big)\Big\}e^{iqx},\\
L_{f}(x)&=\Big\{r_{2}\Theta\Big(-x-\frac{d}{2}\Big)+N\Big[\Theta\Big(x+\frac{d}{2}\Big)\\
&\quad-\Theta\Big(x-\frac{d}{2}\Big)\Big]\Big\}e^{-iqx},
\end{split}
\label{eq4}
\end{equation}
where $q=k-\omega_{f}/v_{g}$ and $\Theta(x)$ is the Heaviside step function. In addition, $A$ and $B$ ($M$ and $N$) are the probability amplitudes of finding a right-moving photon and a left-moving one in the region of $x_{1}<x<x_{2}$, respectively, and the atom finally in state $|g\rangle$ ($|f\rangle$). For the frequency-preserving case (i.e., the final atomic state is $|g\rangle$), we define $t_{1}$ and $r_{1}$ as the transmission and reflection amplitudes of the input photon, respectively, while for the frequency-converting case (i.e., the final atomic state is $|f\rangle$) we define $t_{2}$ and $r_{2}$ as the conversion amplitudes of creating an output photon with wave vectors $q$ and $-q$, respectively.

Substituting Eq.~(\ref{eq4}) into Eq.~(\ref{eq3}), one can obtain
\begin{equation}
\begin{split}
0&=-iv_{g}(A-1)e^{-ikd/2}+g_{1}u_{e},\\
0&=-iv_{g}(t_{1}-A)e^{ikd/2}+g_{1}u_{e},\\
0&=-iv_{g}(r_{1}-B)e^{ikd/2}+g_{1}u_{e},\\
0&=-iv_{g}Be^{-ikd/2}+g_{1}u_{e},\\
0&=-iv_{g}Me^{-iqd/2}+g_{2}u_{e},\\
0&=-iv_{g}(t_{2}-M)e^{iqd/2}+g_{2}u_{e},\\
0&=-iv_{g}(r_{2}-N)e^{iqd/2}+g_{2}u_{e},\\
0&=-iv_{g}Ne^{-iqd/2}+g_{2}u_{e},\\
0&=\frac{g_{1}}{2}[(A+B+1)e^{-ikd/2}+(A+B\\
&\quad\,+t_{1}+r_{1})e^{ikd/2}]+\frac{g_{2}}{2}[(M+N)e^{-iqd/2}\\
&\quad\,+(M+N+t_{2}+r_{2})e^{iqd/2}]-\Delta u_{e},
\end{split}
\label{eq5}
\end{equation}
where $\Delta=E-\omega_{e}=v_{g}k-\omega_{e}$ is the detuning between the input photon and the $|g\rangle\leftrightarrow|e\rangle$ transition. Then the transmission amplitude can be readily obtained as
\begin{equation}
t_{1}=\frac{\Delta-2\Gamma_{1}\sin{\phi_{1}}+2i\Gamma_{2}(1+e^{i\phi_{2}})}{\Delta+2i\Gamma_{1}(1+e^{i\phi_{1}})+2i\Gamma_{2}(1+e^{i\phi_{2}})},
\label{eq6}
\end{equation}
where $\phi_{1}=kd$ and $\phi_{2}=qd$ are the phases accumulated by photons of wave vectors $k$ and $q$ propagating between the two coupling points, respectively. Here $\Gamma_{1}=g_{1}^{2}/v_{g}$ ($\Gamma_{2}=g_{2}^{2}/v_{g}$) is the radiative decay rate from the excited state $|e\rangle$ to the lower-energy state $|g\rangle$ ($|f\rangle$) contributed from each atom-waveguide coupling point. For $d=0$, the transmission amplitude in Eq.~(\ref{eq6}) can be simplified as
\begin{equation}
t_{1}=\frac{\Delta+4i\Gamma_{2}}{\Delta+4i\Gamma_{1}+4i\Gamma_{2}},
\label{eq7}
\end{equation}
which recovers that of a ``small'' $\Lambda$-type atom~\cite{shenFC}. Note that the radiative decay rates are quadrupled here due to the two coupling points. Moreover, we point out that the transmission amplitude becomes
\begin{equation}
t_{1}=\frac{\Delta-2\Gamma_{1}\sin{\phi_{1}}}{\Delta+2i\Gamma_{1}(1+e^{i\phi_{1}})}
\label{eq8}
\end{equation}
in the case of $\Gamma_{2}=0$ (i.e., $g_{2}=0$), which is exactly identical to that of a giant two-level atom~\cite{EITWang}. Similarly, one can obtain the other scattering amplitudes as
\begin{equation}
\begin{split}
&r_{1}=\frac{-i\Gamma_{1}(1+e^{i\phi_{1}})^{2}}{\Delta+2i\Gamma_{1}(1+e^{i\phi_{1}})+2i\Gamma_{2}(1+e^{i\phi_{2}})},\\
&t_{2}=\frac{-i\sqrt{\Gamma_{1}\Gamma_{2}}(1+e^{i\phi_{1}})(1+e^{-i\phi_{2}})}{\Delta+2i\Gamma_{1}(1+e^{i\phi_{1}})+2i\Gamma_{2}(1+e^{i\phi_{2}})},\\
&r_{2}=\frac{-i\sqrt{\Gamma_{1}\Gamma_{2}}(1+e^{i\phi_{1}})(1+e^{i\phi_{2}})}{\Delta+2i\Gamma_{1}(1+e^{i\phi_{1}})+2i\Gamma_{2}(1+e^{i\phi_{2}})}.
\end{split}
\label{eq9}
\end{equation}
Once again, the amplitudes in Eq.~(\ref{eq9}) can be simplified to those of a small $\Lambda$-type atom and those of a giant two-level atom for $d=0$ and $\Gamma_{2}=0$, respectively. It is clear that $|t_{1}|^{2}+|r_{1}|^{2}+|t_{2}|^{2}+|r_{2}|^{2}=1$ due to the energy conservation and $|t_{2}|^{2}\equiv|r_{2}|^{2}$ due to the inherent symmetry. In the following, we define $T_{1}=|t_{1}|^{2}$ and $R_{1}=|r_{1}|^{2}$ as the transmission and reflection rates (without frequency conversion), respectively, and $T_{c}=|t_{2}|^{2}+|r_{2}|^{2}$ as the conversion efficiency. Note that the intrinsic dissipation $\gamma$ of the excited state to the environment (other than the waveguide) can be taken into account via $\Delta\rightarrow\Delta+i\gamma$. In this case the total scattering probability $T_{1}+R_{1}+T_{c}$ should be smaller than unity. However, we do not consider such dissipation in this paper, because it only reduces the scattering probabilities and increases the linewidth of the spectra rather than introducing any qualitative variation~\cite{efficient,shenFC}.

\section{Phase controlled photon scattering}\label{sec3}
\begin{figure*}[ptb]
\centering
\includegraphics[width=17 cm]{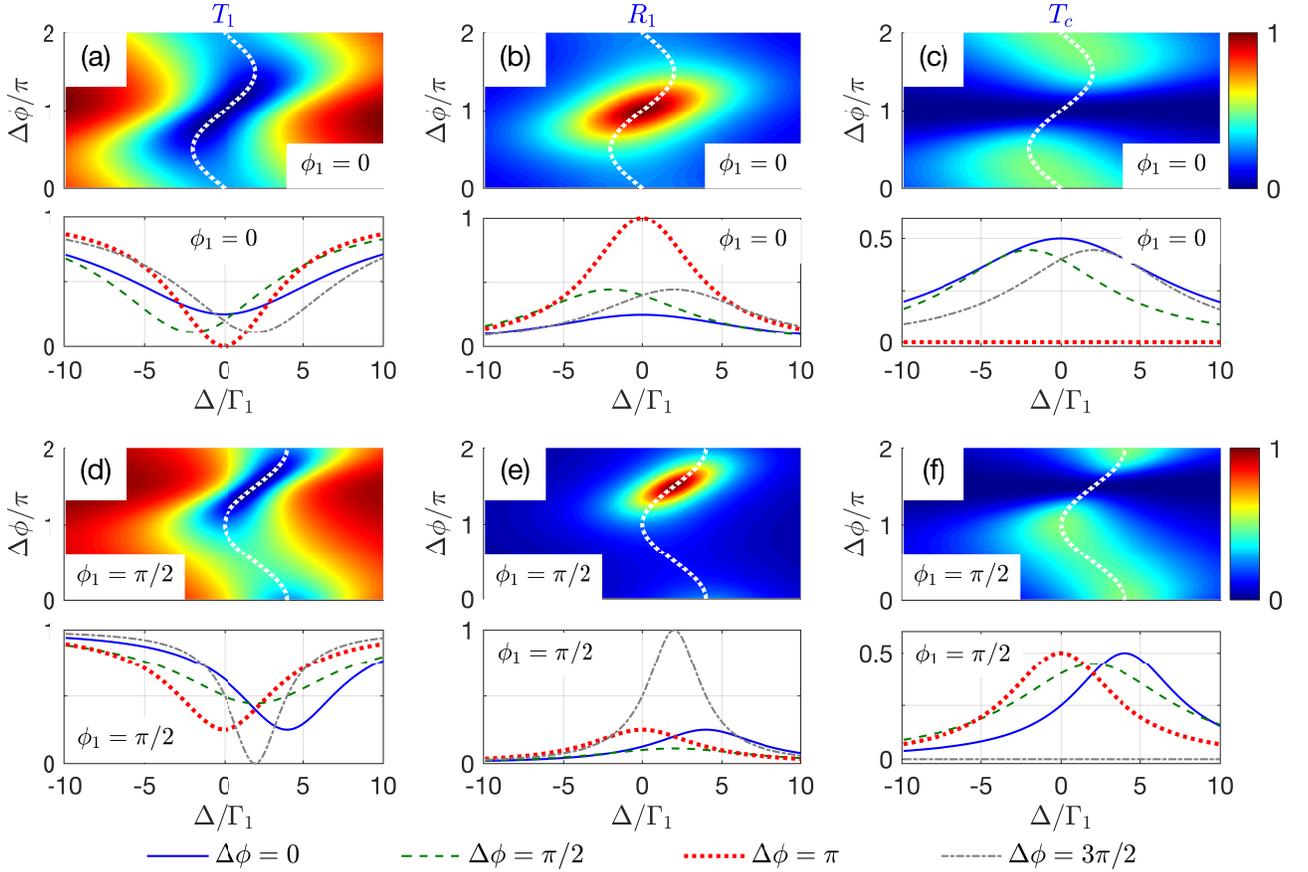}
\caption{Transmission rate $T_{1}$ [(a) and (d)], reflection rate $R_{1}$ [(b) and (e)], and conversion efficiency $T_{c}$ [(c) and (f)] versus detuning $\Delta$ and phase difference $\Delta\phi$ for $\phi_{1}=2m\pi$ (e.g., $\phi_{1}=0$) [(a)-(c)] and $\phi_{1}=(2m+1/2)\pi$ (e.g., $\phi_{1}=\pi/2$) [(d)-(f)]. The white dotted curves in the pseudo-color maps depict the trajectories of $\textrm{min}[T_{1}(\Delta)]$, $\textrm{max}[R_{1}(\Delta)]$, and $\textrm{max}[T_{c}(\Delta)]$ versus $\Delta\phi$. Here we assume $\eta=1$.}\label{fig2}
\end{figure*}

\begin{figure*}[ptb]
\centering
\includegraphics[width=17 cm]{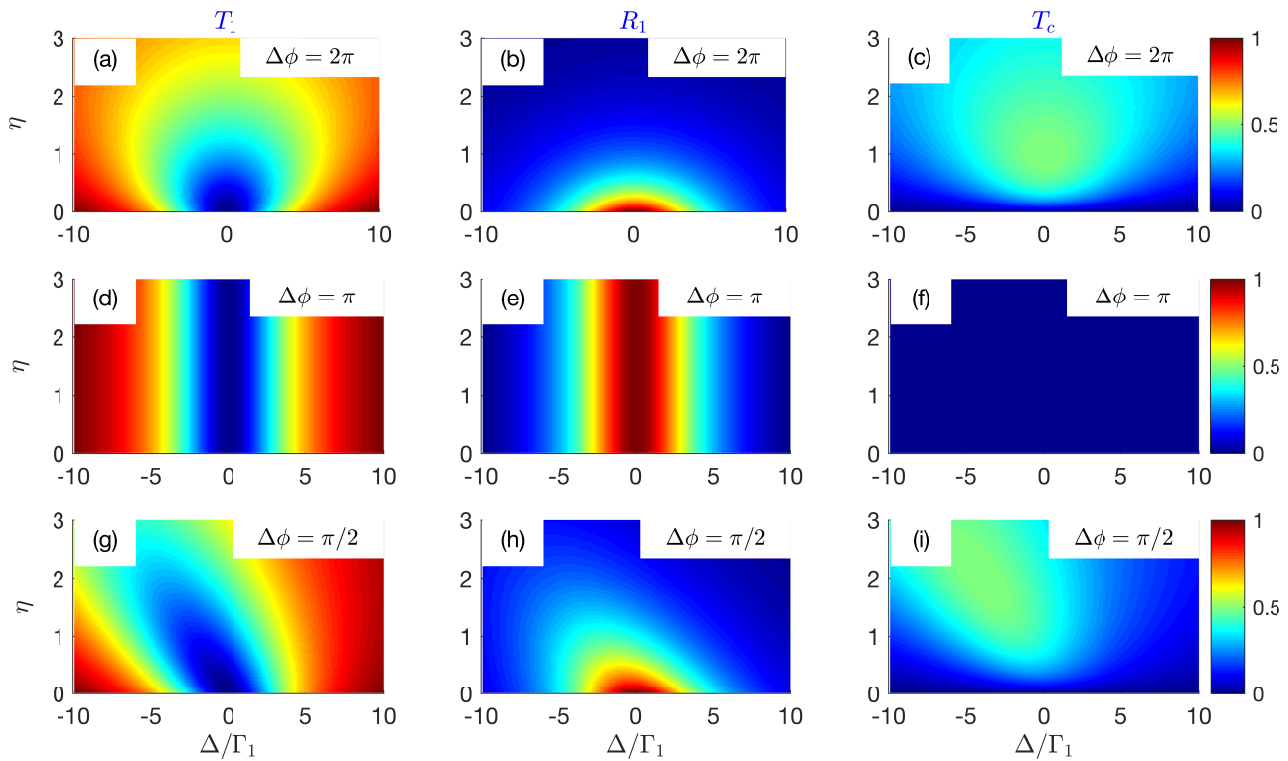}
\caption{Transmission rate $T_{1}$ [(a), (d), (g)], reflection rate $R_{1}$ [(b), (e), (h)], and conversion efficiency $T_{c}$ [(c), (f), (i)] versus detuning $\Delta$ and decay ratio $\eta$ for $\Delta\phi=2\pi$ [(a)-(c)], $\Delta\phi=\pi$ [(d)-(f)], and $\Delta\phi=\pi/2$ [(g)-(i)]. Here we assume $\phi_{1}=2m\pi$ (e.g., $\phi_{1}=0$).}\label{fig3}
\end{figure*}

We first study the dependence of the scattering probabilities on $\phi_{1}$ and $\Delta\phi=\phi_{1}-\phi_{2}$, which both can be tuned experimentally as discussed in detail below. In Fig.~\ref{fig2}, the upper plots of each panel are pseudo-color maps of the scattering probabilities versus detuning $\Delta$ and phase difference $\Delta\phi$, while the lower ones are the two-dimensional profiles for some specific values of $\Delta\phi$. It can be found that both the position and the minimum (maximum) of $T_{1}$ ($R_{1}$ and $T_{c}$) change periodically with $\Delta\phi$. This is reminiscent of the frequency-dependent Lamb shift and decay rate of a giant two-level atom~\cite{Lamb} or a single-mode self-interference resonator~\cite{DLoe,SIDong}. For the giant $\Lambda$-type atom, the effective detuning and linewidth of the spectra are given by the real and imaginary parts of the denominators in Eqs.~(\ref{eq6}) and (\ref{eq9}), respectively, i.e.,
\begin{equation}
\Delta_{\textrm{eff}}=\Delta-2(\Gamma_{1}\sin{\phi_{1}}+\Gamma_{2}\sin{\phi_{2}})
\label{eq10}
\end{equation}
and
\begin{equation}
\Gamma_{\textrm{eff}}=\Gamma_{1,\textrm{eff}}+\Gamma_{2,\textrm{eff}},
\label{eq11}
\end{equation}
with $\Gamma_{1,\textrm{eff}}=2\Gamma_{1}(1+\cos{\phi_{1}})$ [$\Gamma_{2,\textrm{eff}}=2\Gamma_{2}(1+\cos{\phi_{2}})$] the effective decay rate from $|e\rangle$ to $|g\rangle$ ($|f\rangle$) due to the giant-atom structure. Equations~(\ref{eq10}) and (\ref{eq11}) clearly show that the giant-atom interference effects affect the scattering probabilities by modifying the atomic transition frequencies and decay rates. In Figs.~\ref{fig2}(a)-\ref{fig2}(c) we find $T_{1}(\Delta=0)=0$, $R_{1}(\Delta=0)=1$, and $T_{c}(\Delta)\equiv0$ for $\phi_{1}=2m\pi$ and $\Delta\phi=\pi$, implying that total reflection is achieved. In this case, the $|f\rangle\leftrightarrow|e\rangle$ transition is completely suppressed due to the destructive interference of two corresponding decay channels, i.e., $\Gamma_{2}[1+\textrm{exp}(i\phi_{2})]=0$; thus the model reduces to a two-level atom where a resonant incident photon should be perfectly reflected. For the case of $\phi_{1}=(2m+1/2)\pi$, as shown in Figs.~\ref{fig2}(d)-\ref{fig2}(f), the patterns of the pseudo-color maps are shifted along the $y$ axis by $\pi/2$, while the linewidth is reduced by $2\Gamma_{1}$ according to Eq.~(\ref{eq11}). In this case, the condition of total reflection becomes $\Delta\phi=3\pi/2$ and $\Delta=2\Gamma_{1}$. In fact, we have checked that total reflection can always be observed as long as $\phi_{2}=\phi_{1}-\Delta\phi=(2m+1)\pi$ and $\Delta_{\textrm{eff}}=0$. More interestingly, frequency-independent perfect transmission (FIPT), i.e., $T_{1}(\Delta)\equiv1$ and $R_{1}(\Delta)=T_{c}(\Delta)\equiv0$, can be achieved in the case of $\phi_{1}=(2m+1)\pi$, regardless of the value of $\Delta\phi$ (not shown here). This is because the two excitation paths of the $|g\rangle\leftrightarrow|e\rangle$ transition cancel each other completely such that this transition is decoupled from the waveguide. Such a phenomenon has also been revealed with a self-interference resonator, which is referred to as the optical dark states~\cite{DLoe,synphonon}.

It has been shown that the scattering probabilities are determined solely by the decay ratio $\eta=\Gamma_{2}/\Gamma_{1}$ in the small-atom case~\cite{efficient,shenFC} and the optimal frequency conversion occurs when $\eta=1$. For the giant-atom case, however, such a condition becomes phase dependent and thus should be modified. We first demonstrate two special situations, i.e., $\phi_{1}=2m\pi$ and $\Delta\phi=2\pi$ in Figs.~\ref{fig3}(a)-\ref{fig3}(c) and $\phi_{1}=2m\pi$ and $\Delta\phi=\pi$ in Figs.~\ref{fig3}(d)-\ref{fig3}(f). In the former case, the optimal frequency conversion ($T_{c}=0.5$) occurs at $\Delta=0$ and $\eta=1$, which is identical to that of the small-atom case. In the latter case, however, the frequency conversion is completely suppressed and the input photon that is resonant with the $|g\rangle\leftrightarrow|e\rangle$ transition can be totally reflected. In this case, all scattering probabilities are independent of $\eta$. For a more general case, the condition of the optimal conversion can be summarized as $\eta=(1+\cos{\phi_{1}})/(1+\cos{\phi_{2}})$ (i.e., $\eta_{\textrm{eff}}=\Gamma_{2,\textrm{eff}}/\Gamma_{1,\textrm{eff}}=1$) and $\Delta=-2(\Gamma_{1}\sin{\phi_{1}}+\Gamma_{2}\sin{\phi_{2}})$ (i.e., $\Delta_{\textrm{eff}}=0$). For example, as shown in Figs.~\ref{fig3}(g)-\ref{fig3}(i), the optimal frequency conversion occurs at $\eta=2$ and $\Delta=-4\Gamma_{1}$ for $\phi_{1}=2m\pi$ and $\Delta\phi=\pi/2$, which is in good agreement with the analytical condition above.

\section{Efficient frequency conversion with Sagnac interferometers}\label{sec4}
It can be found that the conversion efficiency is still at most $\frac{1}{2}$ in the giant-atom case (see Figs.~\ref{fig2} and \ref{fig3}), although the giant-atom interference effects provide additional tunability for the scattering behaviors. This can be understood from the fact that the giant-atom effects modify the atomic transition frequencies and decay rates such that the giant atom behaves like a small one with renormalized parameters. In this section, we show the feasibility of realizing efficient frequency conversions in the giant-atom case with the assistance of the quantum interferences between counterpropagating modes~\cite{efficient,shenFC,YanSag,zubairy}. In experiments, this can be achieved by coupling the atom with the waveguide through two identical Sagnac interferometers, each of which is connected with the waveguide via a $50:50$ coupler (beam splitter)~\cite{SagExp}, as shown in Fig.~\ref{fig1}(b). At each coupler, the input photon is split equally into two counterpropagating parts (i.e., the clockwise and counterclockwise fields of the Sagnac interferometer), which form photonic superposition states in the interferometers and exhibit quantum interference effects. Moreover, one can also introduce a phase shifter in each Sagnac loop, which can be used for tuning the relative phase $\theta$ between the counterpropagating modes and thus changing their superposition~\cite{exp2}. For superconducting qubits, one can also introduce only one Sagnac loop and couple the qubit with it at two different points~\cite{zubairy}. Supposing that the whole system is symmetric with respect to the spatial inversion, the waveguide modes can be described in terms of the even and odd modes
\begin{equation}
\begin{split}
&a_{e}(x)=\frac{1}{\sqrt{2}}[a_{R}(x)\pm a_{L}(-x)],\\
&a_{o}(x)=\frac{1}{\sqrt{2}}[a_{R}(x)-a_{L}(-x)],
\end{split}
\label{eq12}
\end{equation}
which correspond to $\theta=0$ and $\theta=\pi$ (in the absence of the giant atom), respectively. In this way, $H_{\textrm{w}}$ and $H_{\textrm{int}}$ in Eq.~(\ref{eq1}) can be rewritten as
\begin{equation}
\begin{split}
H_{\textrm{w}}'&=-iv_{g}\int_{-\infty}^{+\infty}dx\Big[a_{e}^{\dag}(x)\frac{\partial}{\partial x}a_{e}(x)+a_{o}^{\dag}(x)\frac{\partial}{\partial x}a_{o}(x)\Big],\\
H_{\textrm{int}}'&=\int_{-\infty}^{+\infty}dxP(x)(\tilde{g}_{1}a_{e}^{\dag}|g\rangle\langle e|+\tilde{g}_{2}a_{e}^{\dag}|f\rangle\langle e|+\text{H.c.}),
\end{split}
\label{eq13}
\end{equation}
where $\tilde{g}_{j}=\sqrt{2}g_{j}$ ($j=1,\,2$). Clearly, the odd mode $a_{o}$ does not interact with the atom and thus can be discarded in the following calculations. In this way, the single-photon state of the system can be given by
\begin{equation}
\begin{split}
|\psi\rangle=&\int_{-\infty}^{+\infty}dx[\psi_{1}(x)a^{\dag}_{e}(x)|0,g\rangle\\
&+\psi_{2}(x)a^{\dag}_{e}(x)|0,f\rangle]+u_{e}|0,e\rangle,
\end{split}
\label{eq14}
\end{equation}
where $\psi_{1}(x)=[R_{g}(x)+L_{g}(-x)]/\sqrt{2}$ and $\psi_{2}(x)=[R_{f}(x)+L_{f}(-x)]/\sqrt{2}$. Then one can obtain
\begin{equation}
\begin{split}
E\psi_{1}(x)&=-iv_{g}\frac{\partial}{\partial x}\psi_{1}(x)+\tilde{g}_{1}P(x)u_{e},\\
E\psi_{2}(x)&=\Big(\omega_{f}-iv_{g}\frac{\partial}{\partial x}\Big)\psi_{2}(x)+\tilde{g}_{2}P(x)u_{e},\\
Eu_{e}&=\omega_{e}u_{e}+P(x)[\tilde{g}_{1}\psi_{1}(x)+\tilde{g}_{2}\psi_{2}(x)]
\end{split}
\label{eq15}
\end{equation}
by solving the stationary Schr\"{o}dinger equation. In this case, the spatial dependence of $\psi_{1}(x)$ and $\psi_{2}(x)$ can be given by
\begin{equation}
\begin{split}
\psi_{1}(x)&=\Big\{\Theta\Big(-x-\frac{d}{2}\Big)+\tilde{A}\Big[\Theta\Big(x+\frac{d}{2}\Big)\\
&\quad-\Theta\Big(x-\frac{d}{2}\Big)\Big]+\tilde{t}_{1}\Theta\Big(x-\frac{d}{2}\Big)\Big\}e^{ikd},\\
\psi_{2}(x)&=\Big\{\tilde{B}\Big[\Theta\Big(x+\frac{d}{2}\Big)-\Theta\Big(x-\frac{d}{2}\Big)\Big]\\
&\quad+\tilde{t}_{2}\Theta\Big(x-\frac{d}{2}\Big)\Big\}e^{iqd}.
\end{split}
\label{eq16}
\end{equation}
Here $\tilde{A}$ and $\tilde{B}$ are the wave amplitudes in the region of $x_{1}<x<x_{2}$ for the frequency-preserving and frequency-converting cases, respectively, while $\tilde{t}_{1}$ and $\tilde{t}_{2}$ denote in this case the transmission (without frequency conversion) and conversion amplitudes, respectively. Combining Eqs.~(\ref{eq15}) and (\ref{eq16}), we have
\begin{equation}
\begin{split}
0&=-iv_{g}(\tilde{A}-1)e^{-i\phi_{1}/2}+\tilde{g}_{1}u_{e},\\
0&=-iv_{g}(\tilde{t}_{1}-\tilde{A})e^{i\phi_{1}/2}+\tilde{g}_{1}u_{e},\\
0&=-iv_{g}\tilde{B}e^{-i\phi_{2}/2}+\tilde{g}_{2}u_{e},\\
0&=-iv_{g}(\tilde{t}_{2}-\tilde{B})e^{i\phi_{2}/2}+\tilde{g}_{2}u_{e},\\
0&=\frac{\tilde{g}_{1}}{2}[(\tilde{A}+1)e^{-i\phi_{1}/2}+(\tilde{A}+\tilde{t}_{1})e^{i\phi_{1}/2}]\\
&\quad\,+\frac{\tilde{g}_{2}}{2}[\tilde{B}e^{-i\phi_{2}/2}+(\tilde{B}+\tilde{t}_{2})e^{i\phi_{2}/2}]-\Delta u_{e},
\end{split}
\label{eq17}
\end{equation}
which results in
\begin{equation}
\begin{split}
&\tilde{t}_{1}=\frac{\Delta-i\tilde{\Gamma}_{1}(1+e^{-i\phi_{1}})+i\tilde{\Gamma}_{2}(1+e^{i\phi_{2}})}{\Delta+i\tilde{\Gamma}_{1}(1+e^{i\phi_{1}})+i\tilde{\Gamma}_{2}(1+e^{i\phi_{2}})},\\
&\tilde{t}_{2}=\frac{-4i\sqrt{\tilde{\Gamma}_{1}\tilde{\Gamma}_{2}}\cos{\frac{\phi_{1}}{2}}\cos{\frac{\phi_{2}}{2}}}{\Delta+i\tilde{\Gamma}_{1}(1+e^{i\phi_{1}})+i\tilde{\Gamma}_{2}(1+e^{i\phi_{2}})}
\end{split}
\label{eq18}
\end{equation}
with $\tilde{\Gamma}_{j}=\tilde{g}_{j}^{2}/v_{g}$. Likewise, $|\tilde{t}_{1}|^{2}+|\tilde{t}_{2}|^{2}=1$ in the case of $\gamma=0$. In the following, we will show that efficient frequency conversion (i.e., $|\tilde{t}_{2}|^{2}=1$) can also be achieved in the giant-atom case by exploiting the Sagnac quantum interferences.

\begin{figure}[ptb]
\centering
\includegraphics[width=8.5 cm]{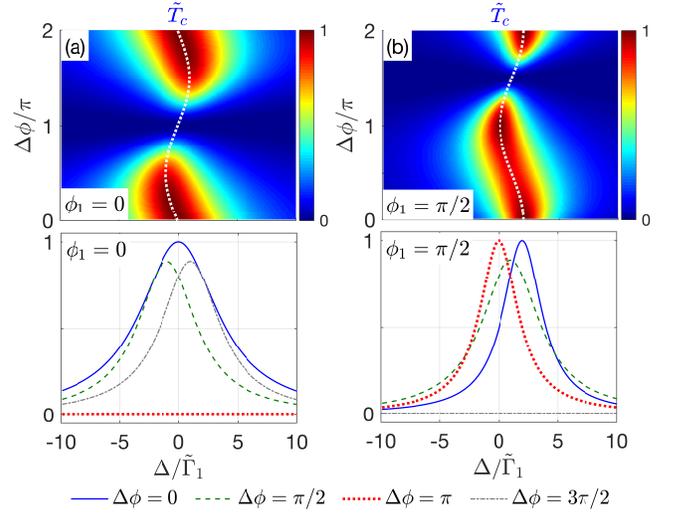}
\caption{Conversion efficiency $\tilde{T}_{c}$ versus detuning $\Delta$ and phase difference $\Delta\phi$ for (a) $\phi_{1}=2m\pi$ (e.g., $\phi_{1}=0$) and (b) $\phi_{1}=(2m+1/2)\pi$ (e.g., $\phi_{1}=\pi/2$). The white dotted curves in the pseudo-color maps depict the trajectories of $\textrm{max}[\tilde{T}_{c}(\Delta)]$ versus $\Delta\phi$. Here we assume $\eta=1$.}\label{fig4}
\end{figure}

We plot in Fig.~\ref{fig4} the conversion efficiency $\tilde{T}_{c}=|\tilde{t}_{2}|^{2}$ versus detuning $\Delta$ and phase difference $\Delta\phi$ for different values of $\phi_{1}$ (similar to Fig.~\ref{fig2}, the upper and lower plots are pseudocolormaps and two-dimensional profiles, respectively). In this case, the transmission rate $\tilde{T}_{1}=|\tilde{t}_{1}|^{2}$ includes both the forward and backward scattering components. It can be simply calculated according to $\tilde{T}_{1}\equiv1-\tilde{T}_{c}$ due to the energy conservation and thus shows inverse patterns with respect to $\tilde{T}_{c}$. Clearly, it can be seen from Fig.~\ref{fig4} that efficient frequency conversion with $\tilde{T}_{c}=1$ can be achieved in this case owing to the quantum interference between the counterpropagating modes in the Sagnac loops. Once again, both the position and the maximum of $\tilde{T}_{c}$ are $\Delta\phi$ dependent with the period of $2\pi$. For $\phi_{1}=2m\pi$, as shown in Fig.~\ref{fig4}(a), the maximum of $\tilde{T}_{c}$ decreases gradually with $\Delta\phi$ until $\tilde{T}_{c}(\Delta)\equiv0$ [i.e., $\tilde{T}_{1}(\Delta)\equiv1$] at $\Delta\phi=\pi$ (this phenomenon is in fact the total reflection shown in Fig.~\ref{fig2}, which is frequency dependent if we divide $\tilde{T}_{1}$ into the forward and backward components). The largest frequency shift with a value of $\tilde{\Gamma}_{2}$ is achieved when $\Delta=\pi/2$. In this case, the effective detuning and linewidth can be given by $\Delta-(\tilde{\Gamma}_{1}\sin{\phi_{1}}+\tilde{\Gamma}_{2}\sin{\phi_{2}})$ and $\tilde{\Gamma}_{1}(1+\cos{\phi_{1}})+\tilde{\Gamma}_{2}(1+\cos{\phi_{2}})$, respectively, which are in fact identical to those in Eqs.~(\ref{eq10}) and (\ref{eq11}) due to $\tilde{\Gamma}_{j}=2\Gamma_{j}$. As shown in Fig.~\ref{fig4}(b), the patterns (see the pseudocolormaps) are shifted along the $y$ axis by $\pi/2$ and the linewidth is reduced by $\tilde{\Gamma}_{1}$ by changing $\phi_{1}$ from $2m\pi$ to $(2m+1/2)\pi$, which is similar to the case without the Sagnac interferences (see Fig.~\ref{fig2}). Note that FIPT is also achievable in this case as long as $\phi_{1}=(2m+1)\pi$, with which $\tilde{T}_{c}(\Delta\phi,\eta)\equiv0$ and $\tilde{T}_{1}(\Delta\phi,\eta)\equiv1$ are independent of all other parameters and input photons that are off-resonance with the $|f\rangle\leftrightarrow|e\rangle$ transition undergo no scattering.

\begin{figure}[ptb]
\centering
\includegraphics[width=8 cm]{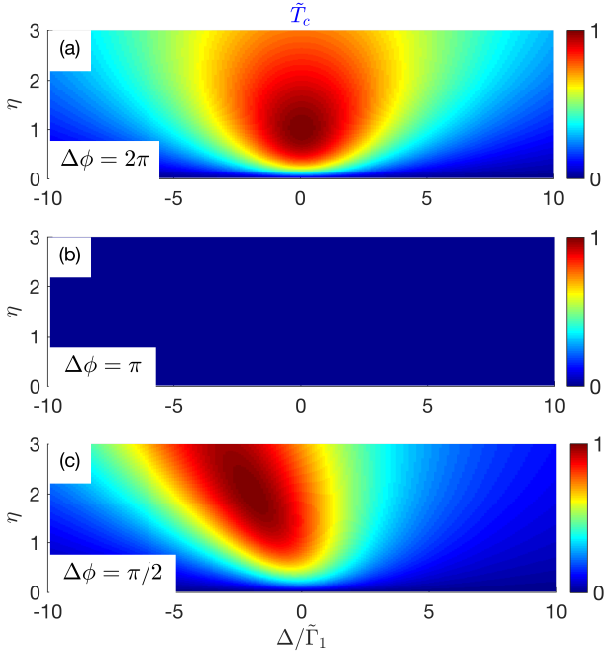}
\caption{Conversion efficiency $\tilde{T}_{c}$ versus detuning $\Delta$ and decay ratio $\eta$ for (a) $\Delta\phi=2\pi$, (b) $\Delta\phi=\pi$, and (c) $\Delta\phi=\pi/2$. Here we assume $\phi_{1}=2m\pi$ (e.g., $\phi_{1}=0$).}\label{fig5}
\end{figure}

Moreover, we demonstrate in Fig.~\ref{fig5} that the condition of the optimal frequency conversion obtained in Sec.~\ref{sec3} still holds in the presence of the Sagnac interferences. Once again, the optimal conversion ($\tilde{T}_{c}=1$ in this case) occurs at $\eta=1$ and $\Delta=0$ if both $\phi_{1}$ and $\Delta\phi$ are integer multiples of $2\pi$, as shown in Fig.~\ref{fig5}(a), while the frequency conversion is completely suppressed over the whole frequency range if $\phi_{2}=\phi_{1}-\Delta\phi=(2m+1)\pi$, as shown in Fig.~\ref{fig5}(b). In addition, the optimal frequency conversion occurs at $\eta=2$ and $\Delta=-2\tilde{\Gamma}_{1}$ in a more general case of $\phi_{1}=2m\pi$ and $\Delta\phi=\pi/2$, as shown in Fig.~\ref{fig5}(c), which is in fact the same condition as that in Figs.~\ref{fig2}(g)-\ref{fig2}(i) due to $\tilde{\Gamma}_{1}=2\Gamma_{1}$.

Finally, we summarize in Table~\ref{tab} the analytical conditions of some aforementioned scattering phenomena. We point out that these conditions are in fact identical for the cases with and without the Sagnac interferences, which implies that the interference effects induced by the giant-atom structure and the Sagnac interferometers are compatible and play their roles independently. We conclude that FIPT or total reflection can be achieved if $\phi_{1}$ or $\phi_{2}$ is an odd multiple of $\pi$, while the optimal frequency conversion demands an appropriate decay ratio $\eta$ for given phases. As a side note, we point out that the intrinsic dissipation $\gamma$ does not change the conditions in Table~\ref{tab}. In other words, such dissipation does not affect the scattering phenomena qualitatively. The only influence is that the scattering probabilities (linewidth) decrease overall (increases) as $\gamma$ increases in the total reflection and optimal frequency conversion cases, whereas the transmission rate remains unity over the whole frequency range in the FIPT case.

\begin{table}
\centering
%\arrayrulewidth=0.5pt
\caption{Conditions of some scattering phenomena.}\label{tab}
\begin{tabular}{l r}
%\multicolumn{2}{|c|}{Conditions for Several Special Phenomena}\\\hline
\hline\hline\\
Phenomena & Conditions \\
\\\hline
Total reflection & $\phi_{1}-\Delta\phi=(2m+1)\pi$ \\
 & [i.e., $\phi_{2}=(2m+1)\pi$] \\
 & $\Delta=2\Gamma_{1}\sin{\phi_{1}}$ \\
 \\
FIPT & $\phi_{1}=(2m+1)\pi$ \\
\\
Optimal frequency & $\eta=(1+\cos{\phi_{1}})/(1+\cos{\phi_{2}})$, \\
conversion & $\Delta=-(\tilde{\Gamma}_{1}\sin{\phi_{1}}+\tilde{\Gamma}_{2}\sin{\phi_{2}})$\\
\hline\hline
\end{tabular}
\end{table}

\section{Experimental implementations of a giant $\Lambda$-type atom}
In this section, we briefly discuss the feasibility of the giant $\Lambda$-type atom considered in this paper. Experimentally, such a model can be achieved by coupling a GaAs quantum dot (which can be confined in a fiber-coupled semiconductor channel waveguide~\cite{confine}) to a U-type bent waveguide (e.g., optical fiber). In this case, $\omega_{e}/2\pi$ and $\omega_{f}/2\pi$ can be tuned at the order of $10^{14}$ and $10^{9}\,\textrm{Hz}$, respectively, depending on the strength of the external magnetic field. For example, one has $\omega_{e}/2\pi=3.7\times10^{14}\,\textrm{Hz}$ and $\omega_{f}/2\pi=6\,\textrm{GHz}$ for the $D^{0}-DX^{0}$ transition where the $g$-factor is measured as $0.44$~\cite{efficient,gfactor}. By tuning the separation $d$ between the two coupling points (i.e., the relative position of the channel waveguide and the fiber) and the strength of the external magnetic field, one can tune the phase difference $\Delta\phi$ within $[0,\,2\pi]$ and the phase $\phi_{1}$ within $[2m,\,2m+2]\pi$ ($m$ is an integer, which is of the order of $10^{5}$ in this case)~\cite{explain}. One can also implement the model with an artificial $\Lambda$-type atom coupled twice with a transmission line, where both $\phi_{1}$ and $\Delta\phi$ can be tuned within $[0,\,2\pi]$ readily by adjusting the external parameters such as the voltages and currents, or the electric and magnetic fields (such that the energy levels can be reconfigured)~\cite{Wreview2}. Note that the non-Markovian retardation effect can be safely neglected in both implementations due to $d/v_{g}\sim10^{-10}\,\textrm{s}\ll1/(\Gamma_{1}+\Gamma_{2})\sim10^{-7}\,\textrm{s}$~\cite{Solano}.

\section{Conclusions}
In summary, we have considered a giant $\Lambda$-type atom which is coupled with a waveguide at two separated points and studied the single-photon scattering at it. A single input photon can either be transmitted or reflected directly without frequency conversion or undergo an inelastic scattering process with converted frequency, depending on which of the two lower-energy states is finally occupied. For the small-atom case, it is known that the scattering behavior is determined only by the ratio of the two waveguide-induced radiative decay rates. For the giant-atom case, however, both elastic and inelastic scattering processes are also dependent on the phase factors which are related to the two transition frequencies as well as the separation between the two coupling points. Similar to a giant two-level atom or a single-mode self-interference resonator, the scattering processes (both elastic and inelastic) of the $\Lambda$-type atom exhibit a phase-dependent frequency shift and linewidth. In particular, each of the two transitions can be completely suppressed when the corresponding coupling channels interfere destructively with each other. The underlying physics can be simply interpreted with the effective frequency shift and linewidth induced by the giant-atom interference effects. In this way, the giant atom is capable of accessing various limits of a small one and thus exhibits a series of limit phenomena such as FIPT and total reflection. To further increase the efficiency of the frequency conversion, we have also introduced quantum interferences between counterpropagating modes by inserting Sagnac interferometers at both coupling points. It was shown that efficient frequency conversion with unity efficiency can be achieved with the assistance of the Sagnac interferences and all the phenomena that arise from the giant-atom interferences can still be observed, which implies that the two kinds of interference effects can play their roles independently. Finally, we have summarized the analytical conditions of some limit phenomena, which have been shown to be identical for the cases with and without the Sagnac interferences.

It is known that a V-type three-level atom with one transition coupled with the waveguide modes and the other one driven by an external field can be effectively described by a $\Lambda$-type energy-level structure in terms of the dressed states~\cite{njpthree,LSzhw}. This implies that our proposal in this paper can be naturally extended to the V-type giant atom, where the decay ratio can be tuned flexibly. Moreover, one can implement a $\Delta$-type giant atom by driving the magnetic dipole transition between the two lower-energy states via a microwave field~\cite{magdipole,jhli}, where phase-dependent nonreciprocal frequency conversion can be expected due to the closed cyclic energy level. We believe that the results in this paper have potential applications in quantum communication and quantum information processing with single photons.

\section*{Acknowledgments}

L.~D. thanks Da-Wu Xiao, Peng~Zhang, Zhihai~Wang, and Yao-Tong Chen for helpful discussions. This work was supported by the Science Challenge Project (Grant No. TZ2018003) and the National Natural Science Foundation of China (Grants No. 11774024, No. 12074030, and No. U1930402).


\begin{thebibliography}{99}
\bibitem {exp2} In this case, the separation $d$ can be tuned by changing the position of the atom along the Sagnac loops (their centers can be coaxial) and tuning the phase shifters to recover the symmetric superposition of the counterpropagating modes~\cite{efficient}.
\bibitem {explain} For $\phi_{1}=kd=w$ and $\Delta\phi=qd=v$ ($w$ and $v$ are arbitrarily chosen values), the separation $d$ and the difference of the wave vector $\Delta k=k-q$ (which is controlled by the external magnetic field) can be determined by simultaneously solving the equations $(k_{0}+\Delta k/2)d=w$ and $\Delta kd=v$, where $k_{0}=(k+q)/2$ is always constant.
\end{thebibliography}
\end{document}